\documentstyle[prc,aps]{revtex}

\begin{document}
\title{Cross Section Limits for the $^{208}$Pb($^{86}$Kr,n)$^{293}$118 Reaction}
\author{K.E. Gregorich$^{1}$, T.N. Ginter$^{1}$, W. Loveland$^{2}$, D. Peterson$^{2}$%
, J.B. Patin$^{1,3}$, C.M. Folden III$^{1,3},$D.C. Hoffman$^{1,3}$, D.M. Lee$%
^{1}$,H. Nitsche$^{1,3}$, J.P. Omtvedt$^{4}$, L.A. Omtvedt$^{4}$,L. Stavsetra%
$^{4},$ R. Sudowe$^{1}$, P.A. Wilk$^{1,3}$, P.M. Zielinski$^{1,3}$ and K.
Aleklett$^{5}$}
\address{$^{1}$Nuclear Science Division, Lawrence Berkeley National\\
Laboratory,Berkeley, CA 94720}
\address{$^{2}$Dept. of Chemistry, Oregon State University, Corvallis,OR 97331}
\address{$^{3}$Department of Chemistry, University of California, Berkeley, CA 94720}
\address{$^{4}$University of Oslo, Oslo, Norway}
\address{$^{5}$Uppsala University, Uppsala, Sweden}
\date{\today }
\maketitle

\begin{abstract}
In April-May, 2001, the previously reported experiment to synthesize element
118 using the $^{208}$Pb($^{86}$Kr,n)$^{293}$118 reaction was repeated. \ \
No events corresponding to the synthesis of element 118 were observed with a
total beam dose of 2.6 x 10$^{18}$ ions. The simple upper limit cross
sections (1 event) were 0.9 and 0.6 pb for evaporation residue magnetic
rigidities of 2.00 $T\,m$ and 2.12 $T\,m$, respectively. \ A more detailed
cross section calculation, accounting for an assumed narrow excitation
function, the energy loss of the beam in traversing the target and the
uncertainty in the magnetic rigidity of the Z=118 recoils is also presented.
Re-analysis of the primary data files from the 1999 experiment showed the
reported element 118 events are not in the original data. The current
results put constraints on the production cross section for synthesis of
very heavy nuclei in cold fusion reactions.
\end{abstract}


\section{Introduction}

In 1999, the synthesis of element 118 (and its decay products) using the $%
^{208}$Pb($^{86}$Kr,n) reaction was \ reported \cite{victor1}. \ This claim
was based on the apparent occurrence of three decay chains, each consisting
of an implanted heavy atom and six subsequent alpha decays, correlated in
time and position. \ A fourth event involving a number of ``escape'' alpha
particles (depositing only part of their energy in the detector) was
reported also \cite{kandv,kandv2}. \ Based upon the above three chains, and
a revised estimate of the beam dose in the 1999 experiments, a cross section
of 7$_{-3}^{+9}$pb was deduced for the synthesis of element 118 at a
projectile energy (center of target, lab system) of 449 MeV. \ \ A new
calibration of the magnetic rigidities (B$\rho $) of the Berkeley Gas-Filled
Separator (BGS) indicates that the reported element 118 compound nucleus
evaporation residues (EVRs) recoiling from the target would have had a B$%
\rho $ of 2.00-$T\,m$ in the 130 Pa He gas of the separator. \ Attempts to
reproduce this result by other groups \cite{gsi,ganil,riken} were
unsuccessful.

Subsequently, the claim to synthesis of element 118 using the $^{208}$Pb($%
^{86}$Kr,n) reaction has been retracted \cite{retract}. \ That retraction
was based upon the absence of the reported \cite{victor1,kandv,kandv2}
chains in a re-analysis of the binary data on the original 1999 data tapes.
An investigation into this matter has concluded that there was scientific
misconduct and data fabrication by one individual\cite{shank}. GSI has also
 reported similar spurious data\cite{spurious}. Despite the retraction of this claim, the current work is
important in establishing definitive upper limits on the production of
element 118 in the $^{208}$Pb($^{86}$Kr,n) reaction.

Motivated in part by the erroneous report [1], there have been a number of
papers \cite{s1,s2,s3,s4,s5,s6,s7,s8,s9,s10,s11,s12,s13} predicting the
structure and decay properties of element 118 and its daughters. \ Should
element 118 be synthesized, it will be interesting to compare these
predictions with the observations. \ In a similar vein, there have been a
number of papers \cite{r1,r2,r3,r4,r5,r6,r7,r8,r9,r10,r11,r12,r13,r14,r15}
dealing with the synthesis of element 118 and the reported production cross
section, which was unexpectedly large. \ Special mention should be made of
the work of Smola\'{n}czuk \cite{r1,r8} which prompted the experimental
measurement [1]. \ Smola\'{n}czuk originally estimated \cite{r1} a
production cross section of 670 pb for the $^{208}$Pb($^{86}$Kr,n) reaction,
an estimate that was later revised \cite{r8} to 5.9 pb. \ Other predictions 
\cite{r10,r11,r12} for the evaporation residue production cross section for
the reaction of 449 MeV $^{86}$Kr with $^{208}$Pb range from 0.005 to 2 pb.
\ These predicted cross sections are generally larger than expected from a
simple logarithmic extrapolation of the trend of cross sections for
reactions of the type $^{208}$Pb(X,n) \cite{gsi} (fig. 1). \ Similar
predictions of a non-exponential decrease in cold fusion cross sections for
reactions leading to elements 116 and lighter elements have been made \cite
{r5}. The physical effect behind this trend was pointed out by Myers and
Swiatecki\cite{r3,r6} as due to a sinking of the Coulomb barrier below the
bombarding energy for symmetric target-projectile combinations, thus
``unshielding'' the saddle point.\ As pointed out by Siwek-Wilczynska and
Wilczynski\cite{r9}, the system in these cases must still evolve from the
dinuclear composite system into the compound nucleus. \ Thus, even the
establishment of upper limits for the $^{208}$Pb($^{86}$Kr,n) cross section
may contribute to our understanding of the large scale collective motion in
very heavy nuclei.

Besides the work reported herein, three recent attempts to reproduce the
observations of [1] have been reported \cite{gsi,ganil,riken}. We summarize
these attempts in table 1 in terms of the beam energy used, the particle
dose and the observed upper limit cross section. All upper limit cross
sections are reported as ``one event upper limits'',{\it \ i.e.}, the cross
section that would have been reported had one event been observed with the
given particle dose, target thickness, separator efficiency, etc. In the
case of the data in \cite{ganil}, a separator efficiency of 50\% \cite
{stodel} was used to calculate the cross section. All excitation energies
were calculated from \cite{r4}.

\section{2001 Experimental Setup}

The reaction $^{208}$Pb($^{86}$Kr,n) was studied at the 88-Inch Cyclotron of
the Lawrence Berkeley National Laboratory, using the Berkeley Gas-filled
Separator \cite{victor}. The experimental apparatus was a modified, improved
version of the apparatus used in [1], including improved detectors and data
acquisition systems, continuous monitoring of the separator gas purity, and
a better monitoring of the $^{86}$Kr beam intensity. A $^{86}$Kr$^{19+}$
beam was accelerated to 457 MeV with an average current of $\sim $1.3 x 10$%
^{12}$ ions/s. The beam went through the 40 $\mu $g/cm$^{2}$ carbon entrance
window of the separator before passing through the $^{208}$Pb target placed
0.5 cm downstream from the window. The targets were 470 $\mu $g/cm$^{2}$
thick (sandwiched between 40 $\mu $g/cm$^{2}$ C on the upstream side and 10 $%
\mu $g/cm$^{2}$ C on the downstream side). Nine of the arc-shaped targets
were mounted on a 35-cm wheel that was rotated at 300 rpm. The beam energy
in the target was 453 - 445 MeV \cite{srim}, encompassing the in-target
energies used in [1]. The beam intensity was monitored by two silicon p-i-n
detectors (mounted at $\pm $ 27 degrees with respect to the incident beam)
that detected elastically scattered beam particles from the target.
Attenuating screens were installed in front of these detectors to reduce the
number of particles reaching them (and any subsequent radiation damage to
the detector). These detectors, the $^{208}$Pb targets and the separator
entrance window were replaced periodically during the run which lasted three
weeks.

The EVRs (E$\sim $131 MeV) were separated spatially in flight from beam
particles and transfer reaction products by their differing magnetic
rigidities in the gas-filled separator. The separator was filled with helium
gas at a pressure of 130 Pa. The expected magnetic rigidities of 131-MeV $%
^{293}$118 EVRs were estimated using the data of Ghiorso {\it et al.} [39].
These estimates were 2.00 $T\,m$ from extrapolation of the data in their
Fig. 3, and 2.10 $T\,m$ from their semi-empirical formula for EVR charge. \
In the current experiments, two settings of the magnetic fields of the
separator were used, 2.00 $T\,m$ and 2.12 $T\,m$. These settings differ by
6\%, the width in magnetic rigidity of the focal plane detector. The
efficiency of the separator for transport and implantation of Z=118 EVRs was
estimated to be $\sim $79 \% using a Monte Carlo simulation described below.

The detector setup at the focal plane of the separator consisted of a
parallel plate avalanche counter (PPAC) \cite{djm} placed $\sim $29 cm
upstream of a Si strip detector. The 10 cm x 10 cm PPAC registered time,
energy loss, and x,y position of the particles passing through it. It has a
thickness equivalent to $\sim $0.6 mg/cm$^{2}$ of carbon. The time-of-flight
of the EVRs between the PPAC and the Si-strip detector was also recorded.
The PPAC was used to distinguish between events arising from beam-related
particles being implanted into the Si-strip detector, and those arising from
the decay of previously implanted atoms. During these experiments, the PPAC
efficiency for detecting beam-related particles depositing between 8 and 14
MeV in the Si-strip detectror was 97.5-99.5\%.

The 300 $\mu $m thick passivated ion implanted silicon strip detector had 32
vertical strips and an active area of 116 mm x 58 mm. The strips were
position sensitive in the vertical (58 mm) direction. The sources used for
position and energy calibrations are summarized in table 2. The energy
resolution of the focal plane detector was measured during the $^{86}$Kr + $%
^{208}$Pb experiments using the 7.45 MeV $^{211}$Po background peak. \ The
energy resolution was 42 keV (FWHM). \ The differences in measured positions
for the $^{252}$No - $^{248}$Fm full energy $\alpha -\alpha $ correlations
in the $^{48}$Ca + $^{206}$Pb study had a Gaussian distribution with a FWHM
of 0.52 mm ($\sigma $ = 0.22 mm). \ \ The measured position resolution for
full energy alpha particles correlated to ``escape'' alpha particles (which
deposited only 0.5 - 3.0 MeV in the detector) was $\sim $1.2 mm (FWHM). \ A
second silicon strip ``punch-through'' detector was installed behind this
detector to reject particles passing through the primary detector. A ``top''
and a ``bottom'' detector were installed upstream of the focal plane
detector to detect escaping alpha particles and fission fragments. The focal
plane detector combined with these ``top'' and ``bottom'' detectors had an
estimated efficiency of 75 \% for the detection of full energy 11 MeV $%
\alpha $-particles following implantation of a $^{293}$118 nucleus.

Any event with E 
\mbox{$>$}%
0.5 MeV in the focal plane Si-strip detector triggered the data acquisition.
Data were recorded in list mode, and included the time of the trigger, the
position and energy signals from the PPAC and the Si-strip detectors, and
energy signals from the ``top'', ``bottom'' and ``punch-through'' detectors.
With the use of buffering ADCs and scalers, the minimum time between
successive events was 15 $\mu $s.

The energies of the $^{293}$118 EVRs, after passing through the PPAC, were
estimated to be $\sim $73 MeV by extrapolation of heavy ion stopping powers,
calculated with SRIM2000 \cite{srim}. \ The $^{293}$118 implantation pulse
height in the Si-strip detector was estimated to be 24-48 MeV after applying
the pulse-height defect determined for Rn EVRs from the 365-MeV $^{86}$Kr + $%
^{120}$Sn reaction.

With a beam current of 1.3 x 10$^{12}$ $^{86}$Kr ions striking the target,
the average total counting rates (E 
\mbox{$>$}%
0.5 MeV) in the focal plane detector were $\sim $40/s and $\sim $10/s at B$%
\rho $ settings of 2.00 and 2.12 $T\,m$, respectively. The average rate of
``alpha particles'' (8-14 MeV with no PPAC signal) was $\sim $0.2/s.

\section{Results}

At separator magnet settings corresponding to EVR magnetic rigidities of
2.00 and 2.12 $T\,m$, the projectile doses were 1.1 x 10$^{18}$ and 1.5 x 10$%
^{18}$, respectively. No event chains similar to those reported in [1] or as
predicted \cite{s1,s2,s3,s4,s5,s6,s7,s8,s9,s10,s11,s12} for the decay of $%
^{293}$118 were observed. Two independent searches were performed for
superheavy element decay sequences. In the first, a search routine in the
GOOSY \cite{GOOSY} environment was used to search for events similar to
those reported in [1], i.e., a search was made for events in which two alpha
particle decays were detected in the focal plane detector within 30 ms and
with positions differing by 
\mbox{$<$}%
1.5 mm. \ The energies of the alpha particles had to be greater than 10 MeV
and they had to be unrelated to beam events (no PPAC signal). \ \ No such $%
\alpha -\alpha $ correlations were found.

A second, less restrictive search was made for EVRs (%
\mbox{$>$}%
20 MeV with a PPAC signal) followed by alpha decays (8-14 MeV in the focal
plane Si-strip detector, no PPAC signal) correlated in position ($\pm $2mm,
same strip) and time (within 10 sec). For this search, one detector strip on
the low-B$\rho $ edge of the detector and three strips on the high-B$\rho $
edge of the detector were excluded because they detected relatively high
rates of scattered beam particles with trajectories bypassing the PPAC. Two
potential EVR-$\alpha $-$\alpha $ chains were identified. Based on the
singles rates for alpha-like events and EVRs, the expected number of random
EVR-$\alpha $-$\alpha $ chains is $\sim $4. Both of the observed EVR-$\alpha 
$-$\alpha $ chains had large differences in the vertical positions for the
parent and daughter alpha particles, 
\mbox{$\vert$}%
$\Delta $p%
\mbox{$\vert$}%
$\geq $ 0.78mm =3.5$\sigma $. Thus, we conclude that there are no valid EVR-$%
\alpha $-$\alpha $ correlations in the data set from this experiment.

One must also consider the possibility that SHEs decay by spontaneous
fission (SF). Spontaneous fission events in which one fragment was detected
in the focal plane detector (E 
\mbox{$>$}%
90 MeV, no PPAC signal) can be confused with scattered $^{86}$Kr beam
particles not vetoed by the PPAC signal. Therefore, we excluded, from the
analysis, three strips on the low-B$\rho $ edge of the detector and four
strips on the high-B$\rho $ edge of the detector to discriminate against
these scattered beam particles. We observed no single fission fragment
signals during the run using these gating conditions,{\it \ i.e.}, no EVR-SF
events.

There were no correlation chains containing at least EVR-$\alpha $-$\alpha $
or EVR-SF with additional full-energy alpha particles (8-14 MeV with no PPAC
signal) or escape alpha particles (with energy deposited in the focal plane
detector 
\mbox{$>$}%
0.5 MeV and no PPAC signal).

The implantation depth in the Si-strip detector for the $^{293}$118 recoils
(E $\sim $73 MeV after passing through the PPAC) was extrapolated from heavy
ion ranges calculated with SRIM2000 \cite{srim}, and is estimated to be 7 $%
\mu $m. From this depth, the efficiency for detecting full energy for
isotropically emitted 8-14 MeV alpha particles is 55\%. The expected decay
sequence \cite{s13,victor1} following implantation of a $^{293}$118 EVR
consists of 6 $\alpha $-particles emitted sequentially within the first 10
seconds after implantation. Using the 55\% $\alpha $-particle efficiency, $%
\varepsilon ,$ together with a binomial series for the probability P, of
observing at least n members of an N-member chain, 
\begin{equation}
P=\sum\limits_{a=n}^{N}\frac{N!}{a!(N-a)!}\varepsilon ^{a}(1-\varepsilon
)^{N-a}
\end{equation}
the probability for detecting at least two full-energy $\alpha $-particles
from a sequence of six is 93\%. Such a binomial treatment can be used to
calculate the efficiency for detection of other postulated decay chains.
Assuming a 100\% event chain detection efficency results in ``one event''
upper limit cross sections of 0.9 pb and 0.6 pb for B$\rho $ settings of
2.00 and 2.12 $T\,m$, respectively. These cross section limits assume
detection of one event, whereas none were observed, and a constant
production cross section throughout the target thickness.

\section{Discussion}

\subsection{Cross Section Calculation}

The standard cross section calculation assumes a constant cross section, $%
\sigma _{const}$, for all beam energies within the target. The number of
events observed, n$_{obs}$, is given by 
\begin{equation}
n_{obs}=\phi t\cdot N_{t}\cdot \sigma _{const}\cdot \varepsilon  \label{seq1}
\end{equation}
where $\phi $t is the integrated beam current, N$_{t}$ is the areal density
of target atoms, and $\varepsilon $ is the experimental efficiency. This
formalism was used in [4-6] for the cross section limits quoted in those
works and used above in Table I and related discussion. \ However, in the
case of the $^{208}$Pb($^{86}$Kr,n)$^{293}$118 reaction, the excitation
function is expected to be narrow and the energy loss of the beam in
traversing the target material is relatively large ($\Delta $E $\sim $ 8
MeV) \cite{srim}, so the assumption of a constant cross section for all beam
energies within the target does not hold. For the purpose of an improved
determination of cross section limits in this experiment, a Gaussian
excitation function, $\sigma $(E), with a full-width at half-maximum (FWHM)
of 5 MeV in the lab frame has been assumed: 
\begin{equation}
\sigma (E)=\sigma _{c}\exp ((E-c)^{2}/2s^{2})  \label{seq2}
\end{equation}
where $\sigma _{c}$ is the cross section at c, the centroid energy, E is the
beam energy at the corresponding target depth, and s is the standard
deviation of the Gaussian (s = 2.12 MeV for a 5-MeV FWHM). This results in
the differential equation 
\begin{equation}
\partial n_{obs}=\phi t\cdot \frac{\partial N_{t}(E)}{\partial E}\cdot
\sigma (E)\cdot \varepsilon (E)\cdot \partial E  \label{seq3}
\end{equation}
where the separator efficiency, $\varepsilon $(E), depends on the beam
energy (depth in target). \ Since the areal number density of target atoms
and the dE/dx of the beam are nearly constant throughout the target, the
areal number density of target atoms within a differential target thickness
element, $\frac{\partial N_{t}(E)}{\partial E}$, is constant and equal to N$%
_{t}$/$\Delta $E, where $\Delta $E is 8.0 MeV for the $^{86}$Kr beam in the $%
^{208}$Pb targets. Integrating over the energy range in the target, and
solving for $\sigma _{c}$ results in 
\begin{equation}
\sigma _{c}=\frac{n_{obs}}{\left[ \phi t\cdot \frac{N_{t}}{\Delta E}\cdot
\int\limits_{E_{i}}^{E_{f}}\varepsilon (E)\exp \left[ -\frac{\left(
E-c\right) ^{2}}{2s^{2}}\right] dE\right] }  \label{seq4}
\end{equation}
where E$_{i}$ and E$_{f}$ are the beam energies at the upstream and
downstream limits of the target. For our experiment, assuming the centroid
of the excitation function corresponds to the center of the target and that
the efficiency is constant throughout the target, $\sigma _{c}$/ $\sigma
_{const}$ = 1.6, and thus the cross section limits presented below are
larger than those calculated in the traditional way by this factor.

\subsection{BGS Efficiency Simulation}

The efficiency of the BGS is limited by the initial position, energy and
angular distributions of recoils exiting the target, and by transmission of
these recoils through the BGS under the influence of the magnetic fields,
and energy loss, multiple scattering and charge exchange in the He fill gas.
All of these effects, together with the effects of the estimated angular and
energy distributions of the $^{86}$Kr beam entering the BGS were calculated
with a Monte Carlo simulation. In this simulation, the initial $^{86}$Kr
beam energies and directions were chosen from an assumed Gaussian energy
distribution (with a centroid of 457 MeV - chosen to put c at the center of
the target, and a 0.3\% FWHM - a typical beam energy width from the 88-Inch
Cyclotron) and Gaussian angular (FWHM=0.9%
${{}^o}$%
- typical for the beamline leading to the BGS) distributions. The beam
energy was corrected for energy loss in the carbon entrance window and
target backings. Points were randomly chosen from the assumed Gaussian
excitation function (c=449 MeV, s=2.12 MeV), and if they were within the
energy range subtended by the target (thickness=0.47 mg/cm$^{2}$, dE/dx=17
MeV/(mg/cm$^{2}$)), the depth of interaction in the target was calculated,
and a simulation of the trajectory of an EVR was initiated. The initial
energy and angle of the EVR were corrected for the effect of isotropic
evaporation of a single neutron. Energy loss and angular scattering in the
remaining target material was calculated for each EVR using SRIM2000\cite
{srim}. After exiting the target, the trajectories through the BGS were
simulated, including the effects of the magnetic fields, charge exchange in
the gas, scattering in the gas, and energy loss in the gas. By comparing the
number of EVRs reaching the focal plane detector in the simulation with the
initial number of $^{86}$Kr beam particles, the effects of the fraction of
the excitation function contained in the target and the BGS efficiency as a
function of target depth were accounted for. \ The ``average'' separator
transport efficiency for the EVRs produced in this reaction was calculated
to be 79\%.

\subsection{Improved $^{293}$118 Cross Section Upper Limits}

Since neither the centroid of the element 118 excitation function nor the
magnetic rigidity of the Z = 118 EVRs are known, numerical simulations of
the BGS efficiency were run using several different choices for the
excitation function centroid energy and average EVR magnetic rigidity. These
simulations gave a set of experimental sensitivities as a function of the
assumed excitation function centroid, and of the assumed magnetic rigidity.
The results from two bombardments, the first with 1.1 x 10$^{18}$ $^{86}$Kr
ions at a BGS magnetic rigidity (B$\rho $ ) of 2.00 $T\,m$ and the second
with 1.5 x 10$^{18}$ Kr ions and B$\rho $ = 2.12 $T\,m$ were combined. The
one-event cross section limits (assuming observation of one element 118
chain where zero events were observed) as a function of assumed $\sigma _{c}$
and B$\rho $ are presented in Fig. 2. The cross section limits reached were
as low as 1.1 pb, and a limit of less than 4.5 pb was reached for compound
nucleus excitation energies from 10.0-16.8 MeV (444.5 
\mbox{$<$}%
E$_{lab}$(MeV) 
\mbox{$<$}%
454.0), covering magnetic rigidities for the recoiling products from
1.94-2.18 $T\,m$.

\section{Summary}

Several experiments have led to one-event cross section upper limits near
0.6 pb for the 449-MeV $^{208}$Pb($^{86}$Kr,n)$^{293}$118 reaction (table
1). \ In fig. 1, we compare this 0.6 pb limit with various theoretical
predictions \cite{r5,r8,r11,r12} for the production of $^{293}$118 in the
reaction of 449-MeV $^{86}$Kr with $^{208}$Pb. This limit is below some of
the predicted values. Combination of sets of upper limit cross sections from
Table 1 results in an upper limit cross section of $\sim $0.2 pb, placing
more stringent limitations on the validity of some of the models. The most
pessimistic estimate \cite{r12} of the evaporation residue cross section
assumes the probability of forming a true compound nucleus, after capture of
the projectile, decreases approximately four orders of magnitude in going
from $^{70}$Zn + $^{208}$Pb to $^{86}$Kr + $^{208}$Pb. This decrease may
counteract any fusion enhancement due to a lowering of the Coulomb barrier
relative to the energy of the fused system in the latter reaction.
Observation of the production of element 118 in the $^{208}$Pb($^{86}$Kr,n)
reaction will require sensitivity to cross sections smaller than $\sim $ 0.2
pb.

\section{Acknowlegments}

We gratefully acknowledge the operations staff of the 88-Inch Cyclotron and
its ion source person D. Wutte for providing intense, steady beams of $^{86}$%
Kr. We thank L.W. Phair for his help in the re-analysis of the 1999 data.\
We thank Patrick Gamman and Helmut Folger for providing the carbon entrance
windows, and the lead targets. Financial support was provided by the Office of High Energy and
Nuclear Physics, Nuclear Physics Division of the U.S. Dept. of Energy, under
contract DE-AC03-76SF00098 and grant DE-FG06-97ER41026, the Norwegian
Research Council (project no. 138507/410) and The Swedish Foundation for
International Cooperation in Research and Higher Education.

%
%

\begin{figure}[tbp]
\caption{The predicted and observed cross sections for the synthesis of
heavy nuclei for cold fusion reactions involving a $^{208}$Pb target.}
\end{figure}

\begin{figure}[tbp]
\caption{The one event upper limit cross sections measured in this work as a
function of assumed excitation function centroid energy and recoil magnetic
rigidity.}
\end{figure}

%
%

\begin{table}[tbp]
\caption{``One event" Upper Limit Cross Sections for Formation of Element
118 in the $^{208}$Pb($^{86}$Kr,n) reaction.}
\label{t1}
\begin{tabular}{ccccc}
E$^{*}$(MeV) & Dose & Separator-B$\rho$ ($T\,m$) & One Event Upper Limit (pb)
& Reference \\ \hline
13.2 & 1.1 x 10$^{18}$ & BGS-2.00 $T\,m$ & 0.9 & This work \\ 
13.2 & 1.5 x 10$^{18}$ & BGS-2.12 $T\,m$ & 0.6 & This work \\ 
13.2 & 1.1 x 10$^{18}$ & GANIL-LISE velocity filter & 2.1 & [5] \\ 
13.2 & 2.0 x 10$^{18}$ & GARIS-2.1 $T\,m$ & 0.6 & [6] \\ 
13.2 & 2.9 x 10$^{18}$ & GSI-SHIP velocity filter & 0.5 & [4] \\ 
15.5 & 0.4 x 10$^{18}$ & GSI-SHIP velocity filter & 3.6 & [4]
\end{tabular}
\end{table}

\begin{table}[tbp]
\caption{BGS Energy Calibration Points}
\label{t2}
\begin{tabular}{ccccc}
Calibration Reaction & Nuclide & Energy(MeV) & Nuclide & Energy (MeV) \\ 
\hline
External source & $^{148}$Gd & 3.183 & $^{239}$Pu & 5.157 \\ 
& $^{241}$Am & 5.486 & $^{244}$Cm & 5.805 \\ 
365-MeV $^{86}$Kr + $^{120}$Sn & $^{204}$Rn & 6.417 & $^{203}$Rn & 6.497,
6.547 \\ 
218-MeV $^{48}$Ca + $^{208}$Pb & $^{254}$No & 8.10 & $^{250}$Fm & 7.43 \\ 
218-MeV $^{48}$Ca + $^{206}$Pb & $^{252}$No & SF & $^{252}$No & 8.42 \\ 
& $^{248}$Fm & 7.87 & $^{244}$Cf & 7.209
\end{tabular}
\end{table}


\begin{references}
\bibitem{victor1}  V. Ninov, {\it et al.}, Phys. Rev. Lett. {\bf 83} 1104
(1999).

\bibitem{kandv}  K.E. Gregorich and V. Ninov, in Origin of the Elements in
the Solar System, O. Manuel, ed. (Kluwer, New York, 2000) pp 21-34.

\bibitem{kandv2}  K.E. Gregorich and V. Ninov, Journal of Nuclear and
Radiochemical Sciences {\bf 1}, 1 (2000).

\bibitem{gsi}  S. Hofmann and G. Munzenberg, Rev. Mod. Phys. {\bf 72}, 733
(2000).

\bibitem{ganil}  C. Stodel, {\it et al.}, AIP Conf. Proc. {\bf 561}, 344
(2001).

\bibitem{riken}  K. Morimoto, AIP Conf. Proc. {\bf 561}, 354 (2001).

\bibitem{retract}  V. Ninov, {\it et al.}, Phys. Rev. Lett {\bf 89}, 039901
(2002).

\bibitem{shank}  Lawrence Berkeley National Laboratory, Internal Report (2002).

\bibitem{spurious}  S. Hofmann, et al. Eur. Phys. J. {\bf A14}, 147 (2002).

\bibitem{s1}  S. \'{C}wiok, W. Nazarewicz, and P.H. Heenen, Phys. Rev. Lett 
{\bf 83}1108 (1999).

\bibitem{s2}  A. Mamdouh, J.M. Pearson, M. Rayet, and F. Tondeur, Nucl.
Phys. {\bf A679}, 337 (2001).

\bibitem{s3}  M. Bender, arXiv:nucl-th/0010094 v1 29 Oct 2000.

\bibitem{s4}  M. Bender, Phys. Rev. C{\bf 61}, 031302 (2000).

\bibitem{s5}  A.T. Kruppa, M. Bender, W. Nazarewicz, P.-G. Reinhard, T.
Vertse, and S. Cwiok, Phys. Rev. C{\bf 61}, 034313 (2000).

\bibitem{s6}  J. Meng and N. Takigawa, Phys. Rev. C{\bf 61}, 064319 (2000).

\bibitem{s7}  Z. Ren and H. Toki, Nucl. Phys. {\bf A689}, 691 (2000).

\bibitem{s8}  M. Bender, W. Nazarewicz, and P.-G. Reinhard,
arXiv:nucl-th/0103065 v1 24 mar 2001.

\bibitem{s9}  Ch. Beckmann, P. Papazoglu, D. Zschiesche, S. Schramm, H.
Stocker, and W. Greiner, arXiv:nucl-th/0106014 v1 6 Jun 2001.

\bibitem{s10}  H. Koura, AIP Conf. Proc.{\bf 561}, 388 (2001).

\bibitem{s11}  P.-G. Reinhard, M. Bender, T. Burvenich, T. Cornelius, P.
Fleischer, and J. A. Maruhn, AIP Conf.Proc. {\bf 561}, 377 (2001).

\bibitem{s12}  G. Royer, J. Phys. G. {\bf 26}, 1149 (2000).

\bibitem{s13}  R. Smola\'{n}czuk, Phys. Rev. C{\bf 56}, 812 (1997).

\bibitem{r1}  R. Smola\'{n}czuk, Phys. Rev. C{\bf 59}, 2634 (1999).

\bibitem{r2}  R. Smola\'{n}czuk, Phys. Rev. Lett. {\bf 83}, 4705 (1999).

\bibitem{r3}  W.D. Myers and W.J. Swiatecki, arXiv:nucl-th/0011075 v1 20 Nov
2000.

\bibitem{r4}  R. Smola\'{n}czuk, Phys. Rev. C{\bf 61}, 011601(1999).

\bibitem{r5}  V. Yu. Denisov and S, Hofmann, Phys. Rev. C{\bf 61}, 034606
(2000).

\bibitem{r6}  W.D. Myers and W.J. Swiatecki, Phys. Rev. C{\bf 62}, 044610
(2000).

\bibitem{r7}  G.G. Adamian, N.V. Antonenko, S.P.Ivanova, and W. Scheid,
Phys. Rev. C{\bf 62}, 064303 (2000).

\bibitem{r8}  R. Smolanczuk, Phys. Rev. C{\bf 63}, 044607 (2001).

\bibitem{r9}  K. Siwek-Wilczynska and J. Wilczynski, Phys. Rev. C{\bf 64},
024611 (2001).

\bibitem{r10}  V. Yu. Denisov, AIP Conf. Proc. {\bf 561}, 433 (2001).

\bibitem{r11}  E. Cherepanov, Pramana {\bf 53}, 619 (1999).

\bibitem{r12}  G.G. Adamian, N.V. Antonenko, A. Diaz-Torres, W. Scheid, and
Yu. M. Tchuvil'sky, AIP Conf. Proc. {\bf 561}, 421 (2001).

\bibitem{r13}  J. Giardina, {\it et al.}, Izv. Akad. Nauk, Ser. Fiz. {\bf 64}%
, 862 (2000).

\bibitem{r14}  R.K. Gupta, M. Balasubramaniam, G.Munzenberg, W. Greiner, and
W. Scheid, J. Phys. G {\bf 27}, 867 (2001).

\bibitem{r15}  I. Muntian, Z. Patyk, and A. Sobiczewski, Acta Phys. Pol. 
{\bf B32}, 691 (2000).

\bibitem{stodel}  C. Stodel, private communication.

\bibitem{victor}  V. Ninov and K.E. Gregorich, ENAM98, B.M. Sherrill, D.J.
Morrissey and C.N. Davids, ed., (AIP, Woodbury, 1999) p. 704.

\bibitem{srim}  J.F. Ziegler, J.P. Biersack and U. Littmark, The Stopping
and Range of Ions in Solids, (Pergamon, New York, 1985); see also
http://www.srim.org.

\bibitem{al}  A. Ghiorso, S. Yashita, M.E. Leino, L. Frank, J. Kalnins, P.
Armbruster, J.-P. Dufour, and P.K. Lemmertz, Nucl. Instru. Meth. {\bf A269},
192 (1988).

\bibitem{djm}  D.\ Swan, J.\ Yurkon, and D.J.\ Morrissey, Nucl.\ Instrum.\
Meth.\ {\bf A348}, 314 (1994).

\bibitem{GOOSY}  See http://www-gsi-vms.gsi.de/anal/home.html.
\end{references}
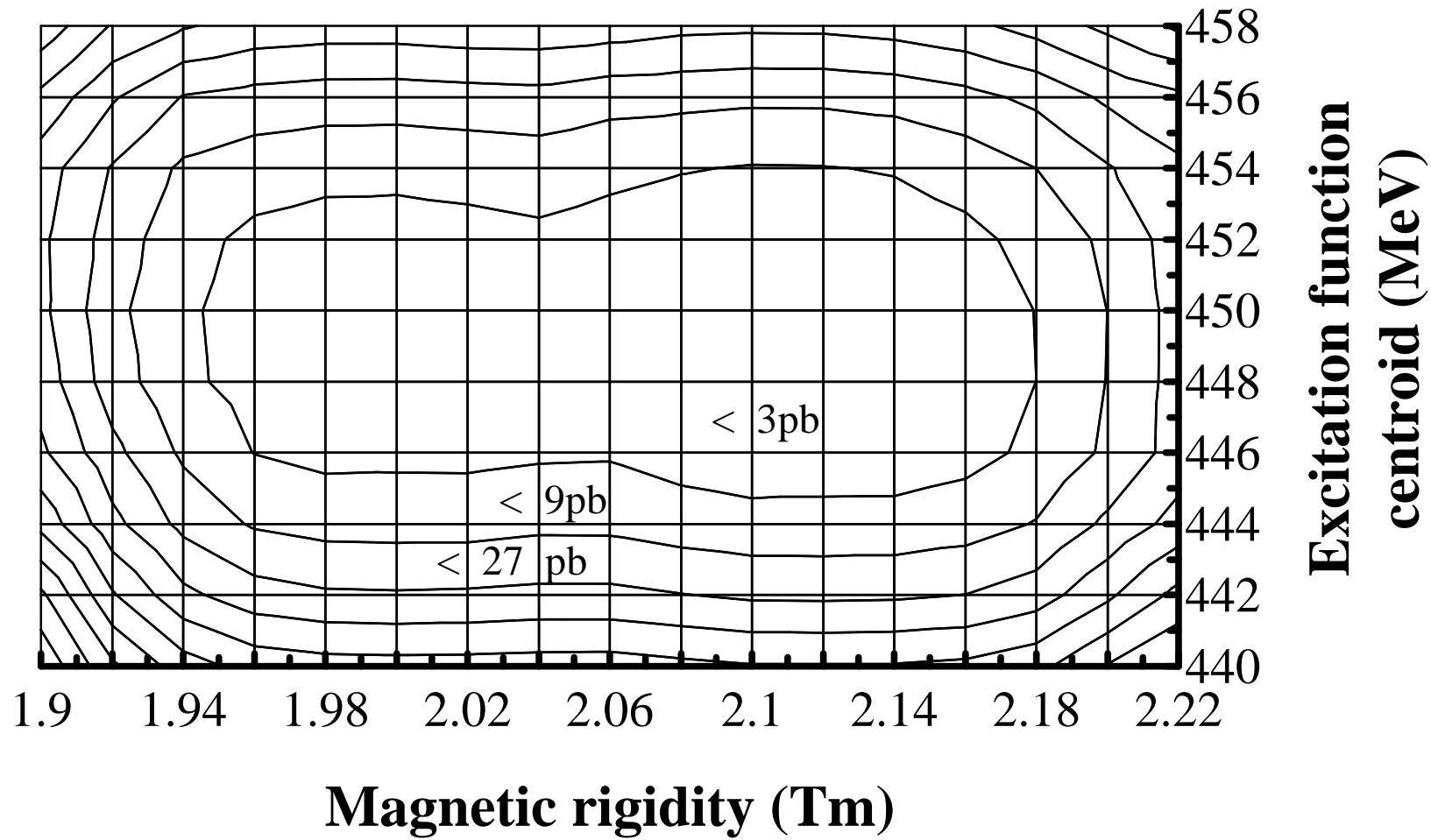
\end{document}